\documentclass[showpacs,preprintnumbers,amsmath,amssymb,twocolumn]{revtex4}
\usepackage{graphicx}
\usepackage{dcolumn}
\usepackage{bm}
\usepackage{amssymb}
\usepackage{array}
\usepackage{hhline}
\usepackage{longtable}

\begin{document}

\title{Influences of degree inhomogeneity on average path length and random walks in disassortative scale-free networks}

\author{Zhongzhi Zhang$^{1,2}$}
\email{zhangzz@fudan.edu.cn}

\author{Yichao Zhang$^{3}$}

\author{Shuigeng Zhou$^{1,2}$}
\email{sgzhou@fudan.edu.cn}

\author{Ming Yin$^{1,2}$}

\author{Jihong Guan$^{3}$}
\email{jhguan@tongji.edu.cn}

\affiliation {$^{1}$School of Computer Science, Fudan University,
Shanghai 200433, China}

\affiliation {$^{2}$Shanghai Key Lab of Intelligent Information
Processing, Fudan University, Shanghai
200433, China} %

\affiliation{$^{3}$Department of Computer Science and Technology,
Tongji University, 4800 Cao'an Road, Shanghai 201804, China}

\begin{abstract}
Various real-life networks exhibit degree correlations and
heterogeneous structure, with the latter being characterized by
power-law degree distribution $P(k)\sim k^{-\gamma}$, where the
degree exponent $\gamma$ describes the extent of heterogeneity. In
this paper, we study analytically the average path length (APL) of
and random walks (RWs) on a family of deterministic networks,
recursive scale-free trees (RSFTs), with negative degree
correlations and various $\gamma \in (2,1+\frac{\ln 3}{\ln 2}]$,
with an aim to explore the impacts of structure heterogeneity on the
APL and RWs. We show that the degree exponent $\gamma$ has no effect
on the APL $d$ of RSFTs: In the full range of $\gamma$, $d$ behaves
as a logarithmic scaling with the number of network nodes $N$ (i.e.
$d \sim \ln N$), which is in sharp contrast to the well-known double
logarithmic scaling ($d \sim \ln \ln N$) previously obtained for
uncorrelated scale-free networks with $2 \leq \gamma <3$. In
addition, we present that some scaling efficiency exponents of
random walks are reliant on the degree exponent $\gamma$.
\end{abstract}

\pacs{89.75.Hc, 89.75.Fb, 05.40.Fb}


\date{\today}
\maketitle

\section{Introduction}

The last decade has witnessed tremendous activities devoted to the
characterization and understanding of real-life systems in nature
and society~\cite{AlBa02,DoMe02,Ne03,BoLaMoChHw06}. Extensive
empirical studies have revealed that most real networked systems
exhibit scale-free behavior~\cite{BaAl99}, which means that these
systems follow a power-law degree distribution $P(k) \sim
k^{-\gamma}$ with degree exponent $\gamma \in [2,3]$. Generally, we
call a network with scale-free behavior a {\em scale-free network}
(SFN), which has a heterogeneous structure encoded in the
characteristic degree exponent $\gamma$: the smaller the $\gamma$,
the stronger the heterogeneity of the network structure. The
inhomogeneous degree distribution of a SFN has a profound effect on
almost all other aspect of the network structure. For example, it
has been established that scale-free behavior is relevant to average
path length (APL)~\cite{WaSt98} in uncorrelated random SFNs, i.e.,
the APL $d(N)$ for a network with node number $N$ depends on
$\gamma$~\cite{CoHa03,ChLu02}: when $\gamma=3$, $d(N) \sim \ln N$;
when $2 \leq \gamma <3$, $d(N) \sim \ln \ln N$.

As known to us all, the ultimate goal of studying network structure
(e.g. degree distribution) is to understand how the dynamical
behaviors are influenced by the underlying topological properties of
the networks~\cite{Ne03,BoLaMoChHw06}. Among many dynamical
processes, a random walk on networks is fundamental to many branches
of science and engineering, and has been the focus of considerable
attention~\cite{NoRi04,SoRebe05,Bobe05,CoBeTeVoKl07,GaSoHaMa07,BaCaPa08,LeYoKi08}.
As a fundamental dynamical process, the random walk is related to
various other dynamics such as transport in media~\cite{HaBe87},
disease spreading~\cite{LlMa01}, target search~\cite{Sh05}, and so
on. On the other hand, the random walk is useful for the study of
network structure, in particular for the average path
length~\cite{NoRi04,LeYoKi08}. It is thus of theoretical and
practical interest to study a random walk on complex networks,
revealing how the structure (e.g. structural heterogeneity) effects
the diffusive behavior of the random walk.

In addition to the scale-free behavior, it has also been observed
that real networks display ubiquitous degree correlations among
nodes~\cite{MsSn02,PaVaVe01,Newman02}, which are usually measured by
two quantities, i.e., average degree of nearest neighbors of nodes
with a given degree~\cite{PaVaVe01} and Pearson correlation
coefficient~\cite{Newman02}, both of which are equivalent to each
other. Degree correlations are important in characterizing network
topology, according to which one can classify complex networks into
categories~\cite{Newman02}: assortative networks, disassortative
networks, and uncorrelated networks. For example, social networks
are usually assortative, while technological and biological networks
disassortative. Furthermore, degree correlations significantly
influence the collective dynamical behaviors, including intentional
attacks on hub nodes~\cite{SoHaMa06,ZhZhZo07}, games~\cite{SzGa07},
and synchronization~\cite{ChHwMaBo06}, to name but a few.

In view of the importance of both the inhomogeneous degree
distribution and degree correlations, some fundamental questions
rise naturally: In heterogenous correlated networks, how does the
structural heterogeneity, characterized by the alterable degree
exponent $\gamma$, affect the scaling character of the average path
length $d(N)$? Does the relation between $d(N)$ and $\gamma$ in
uncorrelated networks also hold for networks with degree
correlations? Is the behavior of random walks related to structural
heterogeneity in correlated networks? Such a series of important
questions still remain open.

In this paper, we study the average path length of and a random walk
on a family of deterministic treelike disassortative scale-free
networks with changeable degree exponent $\gamma \in
\big(2,1+\frac{\ln 3}{\ln 2}\big]$. We choose deterministic networks
as investigation object, because they allow us to study analytically
their topological properties and some dynamic processes running on
them. Our exact results show that in contrast to the scaling
obtained for uncorrelated networks, in their full range of $\gamma$,
the average path length of the considered deterministic networks
grows logarithmically with the number of nodes, and that only
partial scalings of the random walk depend on the degree exponent
$\gamma$.

\begin{figure}
\begin{center}
\includegraphics[width=.75\linewidth,trim=100 20 100 0]{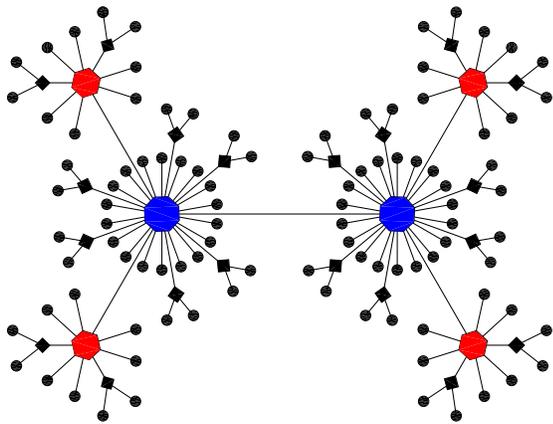} \
\end{center}
\caption[kurzform]{(color online) The first four generations of the
network construction for a special case of $m=2$.} \label{network}
\end{figure}

\section{The recursive scale-free trees}

In this section, we introduce a network model defined in a recursive
way~\cite{JuKiKa02}, which has attracted a great amount of
attention~\cite{GhOhGoKaKi04,CoRobA05,DoMeOl06,ZhZhChGuFaZh07,ZhZhChGu08}.
We call this model {\em recursive scale-free trees}
(RSFTs)~\cite{JuKiKa02}. We investigate RSFTs because of their
intrinsic interest and because these networks have general degree
distribution exponent $\gamma \in \big(2,1+\frac{\ln 3}{\ln
2}\big]$. Moreover, RSFTs are deterministic, which allows us to
study analytically their topological properties and dynamical
processes on them. They are therefore good test-beds and substrate
networks.

The recursive scale-free trees, denoted by $R_{m,t}$ ($m$ is a
positive integer) after $t$ ($t\geq 0$) generation evolution, are
constructed as follows. For $t=0$, $R_{m,0}$ is an edge connecting
two nodes. For $t\geq 1$, $R_{m,t}$ is obtained from $R_{m,t-1}$:
For each of the existing edges in $R_{m,t-1}$, $m$ new nodes are
introduced and connected to either end of the edge.
Figure~\ref{network} shows the construction process for the
particular case of $m=2$.

According to the network construction, one can see that at each step
$t_i$ ($t_i\geq 1$) the number of newly introduced nodes is
$L_v(t_i)=2m(1+2m)^{t_i-1}$. From this result, we can easily compute
network order (i.e., the total number of nodes) $N_t$ at step $t$:
\begin{equation}\label{Nt}
N_t=\sum_{t_i=0}^{t}L_v(t_i)=(2m+1)^{t}+1.
\end{equation}

Let $k_i(t)$ be the degree of a node $i$ at time $t$, which entered
the networks at step $t_i$ ($t_i\geq 0$). Then
\begin{equation}\label{ki}
k_i(t)=(m+1)^{t-t_{i}}.
\end{equation}
From Eq.~(\ref{ki}), one can easily see that at each step the degree
of a node increases $m$ times, i.e.,
\begin{equation}\label{ki2}
k_i(t)=(m+1)\,k_i(t-1).
\end{equation}

RSFTs present some typical characteristics of real-life networks in
nature and society, and their main topological properties are
controlled by the parameter $m$. They have a power law degree
distribution with exponent $\gamma=1+\frac{\ln(2m+1)}{\ln(m+1)}$
belonging to the interval between 2 and
3~\cite{JuKiKa02,ZhZhChGuFaZh07}. The diameter of RSFTs, defined as
the longest shortest distance between any pair of nodes, increases
logarithmically with network order~\cite{ZhZhChGuFaZh07}, that is to
say, RSFTs are small-world. The betweenness distribution exhibits a
power-law behavior with exponent $\gamma_b=2$~\cite{GhOhGoKaKi04}.
In addition, RSFTs are disassortative, the average degree of nearest
neighbors, $k_{\rm nn}(k)$, for nodes with degree $k$ is
approximately a power-law function of $k$ with a negative
exponent~\cite{ZhZhChGuFaZh07}.

After introducing the RSFTs, in what follows we will study the
average path of the recursive scale-free trees and random walks on
them. We will show that the exponent $\gamma$ of degree distribution
has no qualitative effect on APL and mean first-passage time (FPT)
for all nodes, but has essential influence on FPT for old nodes when
the networks grow.

\section{Average path length}

We now study analytically the APL $d_t$ of the recursive scale-free
trees $R_{m,t}$ by using a method similar to but different from that
proposed in Ref.~\cite{HiBe06}. It follows that
\begin{equation}\label{eq:app1}
  d_{t} = \frac{\Phi_t}{N_t(N_t-1)/2}\,,
\end{equation}
where $\Phi_t$ is the total distance between all couples of nodes,
i.e.,
\begin{equation}\label{eq:app2}
  \Phi_t = \sum_{i \in R_{m,t},\, j \in R_{m,t},\, i \neq j} d_{ij},
\end{equation}
in which $d_{ij}$ is the shortest distance between node $i$ and $j$.

\begin{figure}
\begin{center}
\includegraphics[width=.70\linewidth,trim=100 0 100 0]{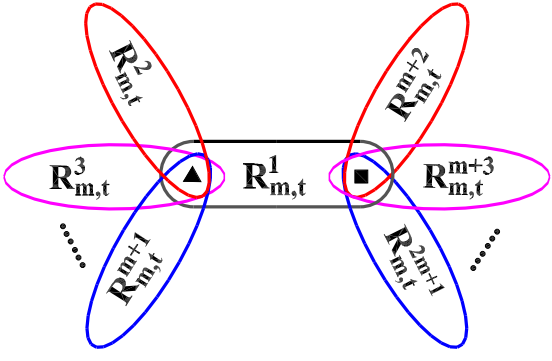}
\caption{(color online) Second construction method of RSFTs. The
graphs after $t+1$ construction steps, $R_{m,t+1}$, may be obtained
by the juxtaposition of $2m+1$ copies of $R_{m,t}$, denoted by
$R_{m,t}^{\theta}$ $(\theta=1,2,3,\cdots,2m, 2m+1)$, which are
connected to one another at the border nodes $X$ ($\blacktriangle$)
and $Y$ ($\blacksquare$).} \label{copy}
\end{center}
\end{figure}

Notice that in addition to the recursive construction, RSFTs can be
alternatively created in another method. Given a generation $t$,
$R_{m,t+1}$ may be obtained by joining at hub nodes $2m+1$ copies of
$R_{m,t}$, see Fig.~\ref{copy}. The second construction method
highlights the self-similarity of RSFTs, which allows us to address
$d_{t}$ analytically. From the obvious self-similar structure, it is
easy to see that the total distance $\Phi_{t+1}$ satisfies the
recursion relation
\begin{equation}\label{eq:app3}
  \Phi_{t+1} = (2m+1)\,\Phi_t + \triangle_t\,,
\end{equation}
where $\triangle_t$ is the sum over all shortest paths whose
endpoints are not in the same $R_{m,t}^{\theta}$ branch. The
solution of Eq.~\eqref{eq:app3} is
\begin{equation}\label{eq:app4}
  \Phi_t = (2m+1)^{t-1} \Phi_1 + \sum_{\tau=1}^{t-1} (2m+1)^{t-\tau-1} \triangle_\tau\,.
\end{equation}
The paths contributing to $\triangle_t$ must all go through at least
either of the two border nodes (i.e., $X$ and $Y$ in
Fig.~\ref{copy}) where the different $R_{m,t}^{\theta}$ branches are
connected. The analytical expression for $\triangle_t$, called the
crossing paths, is found below.

Let $\triangle_t^{\alpha,\beta}$ be the sum of all shortest paths
with endpoints in $R_{m,t}^{\alpha}$ and $R_{m,t}^{\beta}$.
According to whether or not two branches are adjacent, we split the
crossing paths $\triangle_t^{\alpha,\beta}$ into two classes: If
$R_{m,t}^{\alpha}$ and $R_{m,t}^{\beta}$ meet at a border node,
$\triangle_t^{\alpha,\beta}$ rules out the paths where either
endpoint is that shared border node. For example, each path
contributing to $\triangle_t^{1,2}$ should not end at node $X$. If
$R_{m,t}^{\alpha}$ and $R_{m,t}^{\beta}$ do not meet,
$\triangle_t^{\alpha,\beta}$ excludes the paths where either
endpoint is $X$ or $Y$. For instance, each path contributive to
$\triangle_t^{2,m+2}$ should not end at nodes $X$ or $Y$. We can
easily compute that the numbers of the two types of crossing paths
are $m^2+m$ and $m^2$, respectively. On the other hand, any two
crossing paths belonging to the same class have identical length.
Thus, the total sum $\triangle_t$ is given by
\begin{equation}\label{eq:app6}
\triangle_t = (m^2+m)\, \triangle_t^{1,2} + m^2\,
\triangle_t^{2,m+2}\,.
\end{equation}
In order to determine $\triangle_t^{1,2}$ and $\triangle_t^{2,m+2}$,
we define
\begin{align}
\sigma_t = \sum_{i \in R_{m,t},i\ne X}d_{iX}\,. \label{eq:app7}
\end{align}
Considering the self-similar network structure, we can easily know
that at time $t+1$, the quantity $\sigma_{t+1}$ evolves recursively
as
\begin{eqnarray}
\sigma_{t+1}
&=&(m+1)\,\sigma_t+m\left[\sigma_t+(N_t-1)\right]\nonumber\\
&=&(2m+1)\,\sigma_t+m\,(2m+1)^{t}.\label{eq:app8}
\end{eqnarray}
Using $\sigma_0=1$, we have
\begin{eqnarray}
\sigma_t=(mt+2m+1)\,(2m+1)^{t-1}.
\end{eqnarray}
Having obtained $\sigma_t$, the next step is to compute the
quantities $\triangle_t^{1,2}$ and $\triangle_t^{2,m+2}$ given by
\begin{eqnarray}
  \triangle_t^{1,2} &=& \sum_{\substack{i \in R_{m,t}^{1},\,\,j\in
      R_{m,t}^{2}\\ i,j \ne X}} d_{ij}\nonumber\\
  &=& \sum_{\substack{i \in  R_{m,t}^{1},\,\,j\ \in R_{m,t}^{2}\\ i,j \ne X}} (d_{iX} + d_{jX}) \nonumber\\
  &=& (N_t-1)\sum_{\substack{i \in R_{m,t}^{1}\\ i \ne X}} d_{iX} + (N_t-1) \sum_{\substack{j \in R_{m,t}^{2}\\ j \ne X}} d_{jX} \nonumber\\
  &=& 2(N_t-1)\,\sigma_t,
\label{eq:app9}
\end{eqnarray}
and
\begin{eqnarray}
  \triangle_t^{2,m+2} &=& \sum_{\substack{i \in R_{m,t}^{2},\,i \ne X\\ j\in
      R_{m,t}^{m+2},\,j \ne Y}} d_{ij}\nonumber\\
  &=& \sum_{\substack{i \in R_{m,t}^{2},\,i \ne X\\ j\in
      R_{m,t}^{m+2},\,j \ne Y}} (d_{iX} + d_{XY}+ d_{jY}) \nonumber\\
  &=& 2(N_t-1)\,\sigma_t+(N_t-1)^2\,,
\label{eq:app10}
\end{eqnarray}
where $d_{XY}=1$ has been used. Substituting Eqs.~(\ref{eq:app9})
and (\ref{eq:app10}) into Eq.~(\ref{eq:app6}), we obtain
\begin{eqnarray}\label{eq:app12}
\triangle_t &=& 2m(2m+1)(N_t-1)\,\sigma_t+m^2\,(N_t-1)^2\,\nonumber\\
&=&m(2mt+5m+2)(2m+1)^{2t}.
\end{eqnarray}
Inserting Eqs.~(\ref{eq:app12}) for $\triangle_\tau$ into
Eq.~(\ref{eq:app4}), and using $\Phi_1 =5m^2+4m+1$, we have
\begin{equation}
 \Phi_t = \frac{(2m+1)^{t-1}}{2} \left [1+m+(2mt+3m+1)(2m+1)^{t}\right]. \label{eq:app13}
\end{equation}
Plugging Eq.~\eqref{eq:app13} into Eq.~\eqref{eq:app1}, one can
obtain the analytical expression for $d_t$:
\begin{equation}\label{APL}
d_t = \frac {1+m+(2mt+3m+1)(2m+1)^{t}}{(2m+1)\,[(2m+1)^t+1]},
\end{equation}
which approximates $\frac{2mt}{2m+1}$ in the infinite $t$, implying
that the APL shows a logarithmic scaling with network order.
Therefore, RSFTs exhibit a small-world behavior. Notice that this
scaling has been seen previously in some other deterministic
disassortative scale-free networks in the same exponent range, such
as the pseudofractal scale-free web studied
in~\cite{DoGoMe02,ZhZhCh07} and the ``transfractal" recursive
networks addressed in~\cite{RoHaAv07}.

\begin{figure}
\begin{center}
\includegraphics[width=.70\linewidth,trim=120 30 120 30]{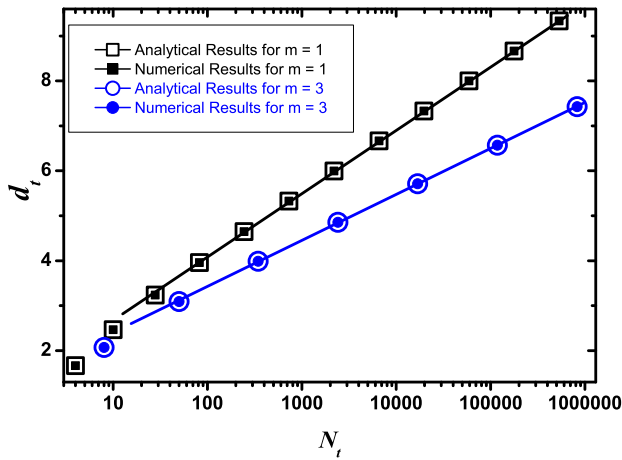}
\end{center}
\caption[kurzform]{\label{distance}(color online) Average path
length $d_{t}$ versus network order $N_{t}$ on a semi-log scale. The
solid lines are guides to the eyes.}
\end{figure}

We have checked our analytic result against numerical calculations
for different $m$ and various $t$. In all the cases we obtain a
complete agreement between our theoretical formula and the results
of numerical investigation, see Fig.~\ref{distance}.

The logarithmic scaling of APL with network order in full rage of
degree exponent $\gamma$ shows that previous relation between APL
and $\gamma$ obtained for uncorrelated scale-free
networks~\cite{CoHa03,ChLu02} is not valid for disassortative
scale-free networks, at least for RSFTs and some other deterministic
scale-free graphs. This leads us to the conclusion that degree
exponent $\gamma$ itself does not suffice to characterize the
average path length of scale-free networks.

\section{Random walks}

This section considers simple random walks on RSFTs defined by a
walker such that at each step the walker, located on a given node,
moves to any of its nearest neighbors with equal probability.

\subsection{Scaling efficiency}

We follow the concept of {\em scaling efficiency} introduced
in~\cite{Bobe05}. Denote by $T_{ij}$ the mean first-passage time
(FPT) between two nodes $i$ and $j$. Let $T_{ii}$ be the mean time
for a walker returning to a node $i$ for the first time after the
walker have left it. When the network order grows from $N$ to $gN$,
one expects that in the infinite limit of $N$
\begin{equation}
T_{ij}(gN)\sim g^{\delta_{ij}}T_{ij}(N),
\end{equation}
where $\delta_{ij}$ is defined as the scaling efficiency exponent.
An analogous relation for $T_{ii}$ defines exponent $\delta_{ii}$.

One can confine scaling efficiency in the nodes already existing in
the networks before growth. Let $T'_{ij}(gN)$ be the mean
first-passage time in the networks under consideration, averaged
over the \emph{original} class of nodes (before growth). Then the
\emph{restricted} scaling efficiency exponent, $\lambda_{ij}$, is
defined by relation
\begin{equation}
T'_{ij}(gN)\sim g^{\lambda_{ij}}T_{ij}(N).
\end{equation}
Similarly, we can define $\lambda_{ii}$.

After introducing the concepts, in the following we will investigate
random walks on RSFTs following a similar method used
in~\cite{Bobe05}. It should be mentioned that our motivation is
different from that of~\cite{Bobe05}. In the work of~\cite{Bobe05},
the authors analyzed scale-free networks with a single degree
distribution exponent $\gamma=1+\frac{\ln3}{\ln2}$, the purpose of
that work is to find what is special about random walks on
scale-free networks, compared to other types of graphs. Here we
study random walks on scale-free networks (RSFTs) with general
$\gamma \in \big(2,1+\frac{\ln3}{\ln2}\big]$. Our aim is to study
the effect of degree exponent $\gamma$ on random walks characterized
by the scaling efficiency proposed in~\cite{Bobe05}.

\subsection{First-passage time for old nodes}

Consider an arbitrary node $i$ in the RSFTs $R_{m,t}$ after $t$
generation evolution. From Eq.~(\ref{ki}), we know that upon growth
of RSFTs to generation $t+1$, the degree $k_i$ of node $i$ increases
$m$ times, namely, from $k_i$ to $(m+1)k_i$.  Let the FPT for going
from node $i$ to any of the $k_i$ old neighbors be $T$. Let the FPT
for going from any of the $mk_i$ new neighbors to one of the $k_i$
old neighbors be $A$. Then we can establish the following equations
\begin{eqnarray}\label{FPT1}
\left\{
\begin{array}{ccc}
T&=&\frac{1}{m+1}+\frac{m}{m+1}(1+A),\\
A&=&1+T,\\
 \end{array}
 \right.
\end{eqnarray}
which leads to $T=2m+1$. Therefore, the passage time from any node
$i$ ($i \in R_{m,t}$) to any node $j$ ($j\in R_{m,t}$) increases
$2m$ times, on average, upon growth of the networks to generation
$t+1$, i.e.,
\begin{equation}\label{FPT2}
 T'_{ij}(N_{t+1})=(2m+1)T_{ij}(N_t).
\end{equation}
Since the network order approximatively grows by $2m$ times in the
large $t$ limit, see Eq.~(\ref{Nt}). This indicates that the scaling
efficiency exponent for old nodes is $\lambda_{ij}=1$, which is a
constant independent of the degree exponent $\gamma$.

Next we continue to consider the return FPT to node $i$. Denote by
$T'_{ii}$ the FPT for returning to node $i$ in $R_{m,t+1}$. Denote
by $T'_{ji}$ the FPT from $j$ --- an old neighbor of $i$ ($i \in
R_{m,t}$) --- to $i$, in $R_{m,t+1}$. Analogously, denote by
$T_{ii}$ the FPT for returning to $i$ in $R_{m,t}$, and $T_{ji}$ the
FPT from the same neighbor $j$, to $i$, in $R_{m,t}$. For $R_{m,t}$,
we have
\begin{equation}\label{FPT3}
 T_{ii}=\frac{1}{k_i}\sum_{j \in \Omega_i(t)}(1+T_{ji})=1+\frac{1}{k_i}\sum_{j \in
 \Omega_i(t)}T_{ji}.
\end{equation}
where $\Omega_i(t)$ is the set of neighbors of node $i$, which
belong to $R_{m,t}$. On the other hand, For $R_{m,t+1}$,
\begin{equation}\label{FPT4}
T'_{ii}=\frac{m}{m+1}\times 2+\frac{1}{m+1}\frac{1}{k_i}\sum_{j \in
\Omega_i(t)}(1+T'_{ji}).
\end{equation}
The first term on the right-hand side (rhs) of Eq.(\ref{FPT4})
accounts for the process in which the walker moves from node $i$ to
the new neighbors and back. Since among all neighbors of node $i$,
$\frac{m}{m+1}$ of them are new, which is obvious from
Eq.~(\ref{ki2}), such a process occurs with a probability
$\frac{m}{m+1}$ and takes two time steps. The second term on the rhs
interprets the process where the walker steps from $i$ to one of the
old neighbors previously existing in $R_{m,t}$ and back, this
process happens with the complimentary probability
$1-\frac{m}{m+1}=\frac{1}{m+1}$.

Using Eq.~(\ref{FPT2}) to simplify Eq.(\ref{FPT4}), we can obtain
\begin{equation}\label{FPT5}
T'_{ii}=\frac{2m+1}{m+1}T_{ii}=(2m+1)^{[1-\ln(m+1)/\ln(2m+1)]}T_{ii}.
\end{equation}
In other words,
\begin{equation}\label{FPT6}
T'_{ii}(N_{t+1})=(2m+1)^{\lambda_{ii}}T_{ii}(N_{t}),
\end{equation}
where the scaling efficiency exponent
$\lambda_{ii}=1-\frac{\ln(m+1)}{\ln(2m+1)}=1-\frac{1}{\gamma-1}$ is
an increasing function of $\gamma$. Thus, the more heterogeneous the
network structure, the more easily for the walker to return to the
origin when the networks grow in size.

\subsection{First-passage time for all nodes}

We now compute $T_{j'j'}$, the FPT to return to a new node $j'\in
R_{m,t}$ that is a neighbor of node $i\in R_{m,t-1}$. Denote by
$T_1$ the FPT from $i$ to $j'$, and $B$ the FPT to return to $i$
(starting off from $i$) without ever visiting $j'$.  Then we have
\begin{equation}\label{FPT7}
T_{j'j'}=1+T_1,
\end{equation}
and
\begin{equation}\label{FPT8}
T_1=\frac{1}{k_i}+\frac{k_i-1}{k_i}(B+T_1).
\end{equation}
Equation~(\ref{FPT8}) can be interpreted as follows: With
probability $\frac{1}{k_i}$ ($k_i$ being the degree of node $i$ in
$R_{m,t}$), the walker starting from node $i$ would take one time
step to go to node $j'$; with the complimentary probability
$\frac{k_i-1}{k_i}$, the walker chooses uniformly a neighbor node
except $j'$, and spends on average time $B$ in returning to $i$,
then takes time $T_1$ to arrives at node $j'$.

In order to close Eqs.~(\ref{FPT7}) and~(\ref{FPT8}), we express the
FPT to return to $i$ as
\begin{equation}\label{FPT9}
T_{ii}(N_{t})=\frac{1}{k_i}\times 2+\frac{k_i-1}{k_i}B.
\end{equation}
Eliminating $T_1$ and $R$, we obtain
\begin{equation}\label{FPT10}
T_{j'j'}(N_{t})=k_iT_{ii}(N_t).
\end{equation}
Combining Eqs.~(\ref{ki2}), (\ref{FPT5}) and~(\ref{FPT10}), we have
\begin{equation}\label{FPT11}
T_{j'j'}(N_{t})=2 (2m+1)^{t}\sim 2N_t.
\end{equation}
Iterating Eqs.~(\ref{FPT5}) and~(\ref{FPT10}), we have that in
$R_{m,t}$ there are $L_v(\epsilon)$ ($0\leq \epsilon \leq t$) nodes
with $T_{ii}=\frac{2(2m+1)^{t}}{(m+1)^{t-\epsilon}}$. Taking average
of $T_{ii}$ over all nodes in $R_{m,t}$ leads to
\begin{eqnarray}\label{FPT12}
\langle
T_{ii}\rangle_t&=&\frac{1}{(2m+1)^{t}+1}\bigg[\frac{4m(2m+1)^{t}}{(m+1)^{t}}+\frac{4m(2m+1)^{t}}{(m+1)^{t-1}}
\nonumber \\&\quad& \quad\quad\quad\quad\quad\quad\quad\times\frac{(2m+1)^{t}(m+1)^{t}-1}{m(2m+3)}\bigg]\nonumber \\
&\overrightarrow{t \to \infty}&\frac{4m+4}{2m+3}(2m+1)^{t} \sim
\frac{4m+4}{2m+3}N_t.
\end{eqnarray}
Equation~(\ref{FPT12}) implies that $\delta_{ii}=1$, which is
uncorrelated with the degree exponent $\gamma$.

We continue to calculate $T_{ij}$ in $R_{m,t}$, which is FPT from an
arbitrary node $i$ to another node $j$. Since each of the newly
created nodes has a degree of 1 and is linked to an old node, the
FPT $T_{i'j}$ from node $i'$ --- a new neighbor of the old node $i$
--- to $j$ equals $T_{ij}$ plus one and thus has little effect on
the scaling when network order $N$ is very large. Therefore, we need
only to consider FPT $T_{ij'}$ from $i$ to $j'$ --- a new neighbor
of $j$, which can be expressed as
\begin{equation}\label{FPT13}
T_{ij'}=T_{ij}+T_{jj'}.
\end{equation}
Notice that
\begin{equation}\label{FPT14}
T_{j'j'}=1+T_{jj'}.
\end{equation}
Substituting Eq.~(\ref{FPT14}) and Eq.~(\ref{FPT11}) for $T_{j'j'}$
into Eq.~(\ref{FPT13}) results in
\begin{equation}\label{FPT15}
T_{ij'}=T_{ij}+2(2m+1)^{t}-1\sim N_t,
\end{equation}
where Eq.~(\ref{FPT2}) has been used. Therefore, we have
\begin{equation}\label{FPT16}
\langle T_{ij}\rangle_t\sim N_t,
\end{equation}
which shows that mean transit time between arbitrary pair of nodes
is proportional to network order. Equation~(\ref{FPT16}) also
reveals that $\delta_{ij}$ is a constant 1, which does not depend on
$\gamma$.

\section{Conclusions}

To explore the effect of structural heterogeneity on the scalings of
average path length and random walks occurring on disassortative
scale-free networks, we have studied analytically a class
deterministic scale-free networks---recursive scale-free trees
(RSFTs)---with various degree exponent $\gamma$. In addition to
scale-free distribution, RSFTs also reproduce some other remarkable
properties of many natural and man-made networks: small average path
length, power-law distribution of betweenness distribution, and
negative degree correlations. They can thus mimic some real systems
to some extent.

With the help of recursion relations derived from the self-similar
structure, we have obtained the solution of average path length for
RSFTs. In contrast to the well-known result that for uncorrelated
scale-free networks with network order $N$ and degree exponent $2
\leq \gamma <3$, their average path length $d(N)$ behaves as a
double logarithmic scaling, $d(N) \sim \ln \ln N$, our rigorous
solution shows that the APL of RSFTs behaves as a logarithmic
scaling, in despite of their degree exponent $\gamma \in
\big(2,1+\frac{\ln 3}{\ln 2}\big]$. Therefore, degree correlations
have a profound impact on the average path length of scale-free
networks.

We have also investigated analytically random walks on RSFTs. We
have shown that for the full range of $\gamma$, the mean transit
time $T_{ij}(N)$ between two nodes averaged over all node pairs
grows linearly with network order $N$. The same scaling holds for
the FPT $T_{ii}(N)$ for returning back to the origin $i$ after the
walker has started from $i$. Thus, despite different $\gamma$, all
the RSFTs exhibit identical scalings of FPT and return FPT for all
nodes. On the other hand, for those nodes already existing in the
networks before growth, the restricted scaling efficiency exponents
are $\lambda_{ij}=1$ and $\lambda_{ii}=1-\frac{1}{\gamma-1}$, where
$\lambda_{ij}$ is not pertinent to $\gamma$, but $\lambda_{ii}$
depends on $\gamma$.

We should stress that our conclusions were drawn only from a
particular type of deterministic treelike disassortative networks.
It is still unknown whether the conclusions are also valid for
stochastic disassortative networks, especially for networks in the
presence of loops. But our results may provide some insights into
random walk problem on complex networks, in particular on trees.
More recently, the so-called border tree motifs have been shown to
be significantly present in real networks~\cite{ViroTrCo08}, looking
from this angle, our work may also shed light on some real-world
systems. Finally, we believe that our analytical techniques could be
helpful for computing average path length of and transit time for
random walks on other deterministic media. Moreover, since exact
solutions can serve for a guide to and a test of approximate
solutions or numerical simulations, we also believe that our
vigorous closed-form solutions can prompt related studies of random
networks.

\subsection*{Acknowledgment}
This research was supported by the National Basic Research Program
of China under grant No. 2007CB310806, the National Natural Science
Foundation of China under Grant Nos. 60704044, 60873040 and
60873070, Shanghai Leading Academic Discipline Project No. B114, and
the Program for New Century Excellent Talents in University of China
(NCET-06-0376).

\end{document}